\documentclass[aps,superscriptaddress,twocolumn]{revtex4}
\usepackage{graphicx}
\usepackage{amsmath}
\usepackage{color}
\usepackage{soul}
\usepackage{placeins}

\begin{document}
 
\title{Deriving models for the Kitaev spin-liquid candidate material $\alpha$-RuCl$_3$ from first principles}
 
\author{Casey Eichstaedt}
\affiliation{Department of Physics \& Astronomy, The University of Tennessee, Knoxville, TN 37996, USA }
\affiliation{Joint Institute of Advanced Materials, The University of Tennessee, Knoxville, Tennessee 37996, USA}
\author{Yi Zhang}
\affiliation{Department of Physics \& Astronomy, Louisiana State University, Baton Rouge, Louisiana 70803, USA}
\affiliation{Center for Computation \& Technology, Louisiana State University, Baton Rouge, Louisiana 70803, USA}
\author{Pontus Laurell}
\affiliation{Center for Nanophase Materials Sciences, Oak Ridge National Laboratory, Oak Ridge, TN 37831, USA}
\author{Satoshi Okamoto} 
\email{okapon@ornl.gov}
\affiliation{Materials Science and Technology Division, Oak Ridge National Laboratory, Oak Ridge, Tennessee 37831, USA}
\author{Adolfo G. Eguiluz}
\email{eguiluz@utk.edu}
\affiliation{Department of Physics \& Astronomy, The University of Tennessee, Knoxville, TN 37996, USA }
\affiliation{Joint Institute of Advanced Materials, The University of Tennessee, Knoxville, Tennessee 37996, USA}
\author{Tom Berlijn}
\email{tberlijn@gmail.com}
\affiliation{Center for Nanophase Materials Sciences, Oak Ridge National Laboratory, Oak Ridge, TN 37831, USA}
\affiliation{Computational Sciences and Engineering Division, Oak Ridge National Laboratory, Oak Ridge, Tennessee 37831, USA}

\begin{abstract}
We use the constrained random phase approximation (cRPA) to derive from first principles the Ru-$t_{2g}$ Wannier function based model for the Kitaev spin-liquid candidate material $\alpha$-RuCl$_3$. We find the non-local Coulomb repulsion to be sizable compared to the local one. In addition we obtain the contribution to the Hamiltonian from the spin-orbit coupling and find it to also contain non-negligible non-local terms. We invoke strong coupling perturbation theory to investigate the influence of these non-local elements of the Coulomb repulsion and the spin-orbit coupling on the magnetic interactions. We find that the non-local Coulomb repulsions cause a strong enhancement of the magnetic interactions, which deviate from experimental fits reported in the literature. Our results contribute to the understanding and design of quantum spin liquid materials via first principles calculations.
\end{abstract}

\maketitle

\section{Introduction}

In his seminal paper, Alexei Kitaev presented an exact solution of the Kitaev model and found it to host a quantum spin-liquid ground state with fractionalized Majorana fermion and gauge flux excitations.~\cite{kitaev_2006} This exotic state of matter is not only interesting from a fundamental scientific point of view but also has been proposed to have potential applications in topological quantum computing.~\cite{kitaev_2003,nayak_2008_rmp} Further progress was made by the idea that the Kitaev quantum spin liquid can possibly be realized in the materials family of the honeycomb iridates A$_2$IrO$_3$ with A=Na,Li.~\cite{jackeli_2009, chaloupka_2010} Assuming that in A$_2$IrO$_3$ the electrons are in the strong coupling limit, in which the interactions dominate over the kinetic energy, and taking into account the spin-orbit coupling, oxidation state and crystal field splitting in the Ir atoms, it was concluded that this compound contains strong Kitaev interactions, in addition to the usual Heisenberg exchange couplings. Depending on the materials parameters it was found that the system can be pushed from an antiferromagnetic (AFM) stripy state into the desired quantum spin liquid state. However, based on combined theoretical and experimental findings it was deduced that A$_2$IrO$_3$ displays AFM zigzag order instead of the AFM stripy order or the quantum spin liquid ground state.~\cite{liu_2011} This was later confirmed by other experiments.~\cite{choi_2012,feng_2012} To account for the experimentally observed zigzag state it was clear that an accurate description of A$_2$IrO$_3$ needed to involve extension beyond the Heisenberg-Kitaev model. To that end it was proposed that second and third nearest neighbor Heisenberg exchange couplings can stabilize the experimentally observed AFM zigzag configuration.~\cite{kimchi_2011} Alternatively, first principles simulations have shown that A$_2$IrO$_3$ contains strong nearest neighbor magnetic anisotropic interactions that favor the AFM zigzag state.~\cite{yamaji_2014_prl} In a third opposite picture it is assumed that A$_2$IrO$_3$ is not in the strong coupling limit, but that instead the strong oxygen assisted hopping between the Ir atoms causes the electrons to delocalize into quasi-molecular orbitals.~\cite{mazin_2012}
   
Another closely related Kitaev spin-liquid candidate material is $\alpha$-RuCl$_3$. The chemically active Ru transition metals in this compound form a honeycomb lattice with five $d$ electrons per atom with strong spin-orbit coupling and electron-electron interactions in the presence of an octahedral crystal field induced by the Cl anions. Therefore, just like with A$_2$IrO$_3$, the materials specifics of $\alpha$-RuCl$_3$ appear to fulfill the conditions laid out in Ref.~\cite{chaloupka_2010} for the emergence of Kitaev interactions.~\cite{plumb_2014,kim_2015}  Inelastic neutron scattering experiments~\cite{banerjee_2016_nmat} on $\alpha$-RuCl$_3$  displayed in addition to AFM zig-zag order~\cite{sears_2015,johnson_2015,banerjee_2016_nmat,cao_2016} a broad continuum in the magnetic excitation spectrum that is indicative of fractionalized excitations. This led to the conclusion that $\alpha$-RuCl$_3$ is proximate to being in the desired quantum spin-liquid phase.~\cite{banerjee_2016_nmat} More recent neutron scattering experiments have shown that the AFM zig-zag order can be suppressed by applying an 8 T magnetic field yielding a magnetic excitation spectrum consistent with a quantum spin liquid phase.~\cite{banerjee_2018} Further evidence for the quantum spin liquid phase has been provided by the observation of the thermal quantum Hall effect in $\alpha$-RuCl$_3$ at similar magnetic field strengths.~\cite{kasahara_2018_nat} 

In order to understand the properties of $\alpha$-RuCl$_3$  and to investigate how this material can be manipulated towards potential applications in topological quantum  computing a microscopic understanding is essential. For that purpose there has been a large effort to map out the magnetic exchange couplings of $\alpha$-RuCl$_3$ both via experimental and theoretical techniques.~\cite{winter_2017,wu_2018, lampenkelley_2018,kim_2016,yadav_2016, winter_2016,suzuki_2018, winter_2017_jpcm,ran_2017,wei_2017} On the experimental side models have been derived by fitting a generalized spin model to various experiments such as inelastic neutron scattering~\cite{winter_2017, ran_2017}, THz spectroscopy~\cite{wu_2018}, anisotropic susceptibility measurements~\cite{lampenkelley_2018}, magnetic specific heat measurements~\cite{suzuki_2018,kubota_2015,do_2017}, and thermal Hall effect measurements~\cite{cookmeyer_2018,kasahara_2018}. The derived magnetic interaction via these fits however display large variations depending on the experiments. In some of the purely theoretical approaches the magnetic interactions are derived by computing  the hopping parameters of the Ru-$t_{2g}$ electrons from first principles while their interaction parameters are taken to be some assumed values.~\cite{kim_2015,kim_2016,winter_2016,hou_2017,wei_2017} In another theoretical approach the first neighboring magnetic interactions are derived from first principles via quantum chemistry techniques, while the second and third neighboring magnetic exchange couplings are modeled phenomenologically.~\cite{yadav_2016} However, there has not been to the best of our knowledge an attempt to derive the magnetic model of $\alpha$-RuCl$_3$ fully from first principles.    

In this paper, we derive the spin model of $\alpha$-RuCl$_3$ fully from first principles. To that end, we first employ Density-Functional Theory (DFT), the constrained Random Phase Approximation (cRPA), and the projected Wannier function method, to obtain a low-energy, generalized Hubbard Hamiltonian for a Hilbert space spanned by Ru-$t_{2g}$ Wannier orbitals.  In a second stage we apply second-order perturbation theory in the strong-coupling limit to our Hamiltonian and obtain the model with spin-degrees of freedom only. We find that in the generalized Hubbard model the inter-atomic Coulomb repulsions and spin-orbit coupling effects are relatively strong compared to their intra-atomic counter-parts. The effect of the inter-atomic interactions is found to strongly enhance the nearest neighboring magnetic couplings by a factor 3-7. The effects of the inter-atomic spin-orbit effects is mainly to enhance the Kitaev coupling by 14\%. The magnetic interactions in our first-principles spin model deviate significantly from the values obtained by fitting experiments. We discuss potential shortcomings in our theoretical approach. Our findings allow for a better understanding of $\alpha$-RuCl$_3$ and quantum spin liquid materials in general via first-principles calculations.    

\section{Methods}

\begin{figure}[h!]
\includegraphics[width=1.0\columnwidth]{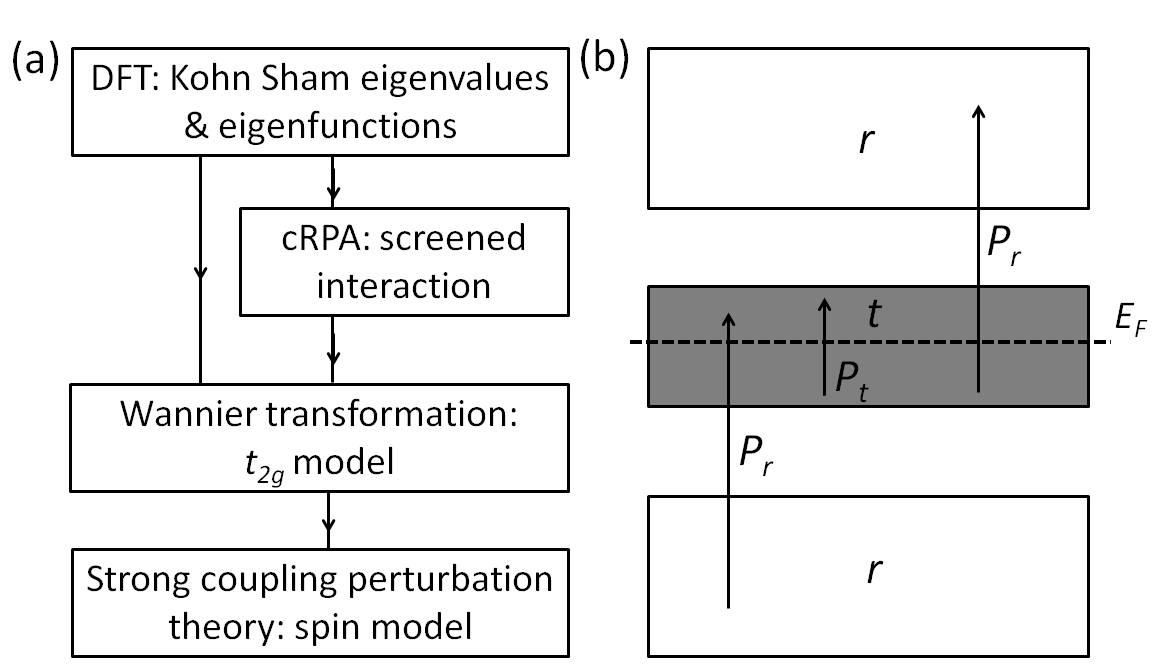}
 \caption{(a) Flow diagram of methods used: Density Functional Theory (DFT), constrained Random Phase Approximation (cRPA), Wannier functions and strong coupling perturbation theory. (b) Screening processes excluded/included in the cRPA denoted by $P_t$/$P_r$, with $t$ labeling the bands in the target space (i.e. the Ru-$t_{2g}$ bands) and $r$ labeling the rest of the bands (adapted from Ref. ~\cite{aryasetiawan_2004}).}
 \label{fig:fig1}
\end{figure}

In order to derive the spin-model for $\alpha$-RuCl$_3$ from first principles we perform the four step procedure described in Fig. \ref{fig:fig1}(a). Below we will briefly review the methodology behind each of the steps. More details about these methods can be found in Ref.~\cite{aryasetiawan_2004,kozhevnikov_2010,yamaji_2014_prl,arita_2012}.  
 
The first step is to perform Density Functional Theory (DFT) calculations of $\alpha$-RuCl$_3$ to obtain the Kohn-Sham eigenvalues and eigenfunctions. The DFT calculations were performed within the generalized gradient approximation using the Perdew-Burke-Ernzerhof exchange correlation scheme~\cite{pbe_1996} and the Linear Augmented Plane Wave (LAPW) method as implemented in the Elk code.~\cite{elk} The space group $C2/m$ and the structural parameters of $\alpha$-RuCl$_3$ are taken from neutron diffraction.~\cite{johnson_2015}. The DFT with and without spin-orbit coupling is performed in the second variational treatment and the relativistic scalar approximation respectively.~\cite{singh_2006}  To compute the interaction matrices (defined in Eq. (\ref{eq:Umat}) and (\ref{eq:Jmat}) below) we include 60 states above the Fermi level and use 2213 LAPW basis functions for the local interactions, and 701 LAPW basis functions for the non-local interactions. The calculations of the hopping parameters and the interaction matrices are performed on a $13\times13\times 7$ and a $6\times6\times 4$  $k$-grid respectively.

The second step is to derive the effective electron-electron interactions using the constrained Random Phase Approximation (cRPA)~\cite{aryasetiawan_2004}. To that end the Hilbert space is divided into two subspaces, the target space $t$ consisting of the bands close to the Fermi level bands and the ``rest'' space $r$ consisting of all the other bands (see Fig.\ref{fig:fig1}(b)). Within the cRPA the effective interaction $W_r(x,x',\omega)$ of the states in the target space $t$ screened by the states in the ``rest'' space $r$ is  
\begin{eqnarray}
W_r(x,x',\omega)=
\nonumber \\
\int d^3y \int d^3y' (1-v(x,y)P_r(y,y',\omega))^{-1}v(y',x') , 
\end{eqnarray}
with $v(x,x')$ the bare Coulomb repulsion and $P_r(x,x',\omega)$ the constrained polarization given by:
\begin{eqnarray}\label{eqPr}
P_r(x,x',\omega)&=&\sum_{k,j}^{occ}\sum_{k',j'}^{unocc}
\Big(
\frac{\langle kj|x\rangle\langle x|k'j'\rangle\langle k'j'|x'\rangle\langle x'|kj\rangle}
{\omega-\epsilon_{k'j'}+\epsilon_{kj}+i0^{+}}
\nonumber \\ &&
-\frac{\langle k'j'|x\rangle\langle x|kj\rangle\langle kj|x'\rangle\langle x'|k'j'\rangle}
{\omega+\epsilon_{k'j'}-\epsilon_{kj}-i0^{+}}
\Big)
\end{eqnarray}
with $\epsilon_{kj}$ and $\langle x| kj\rangle$ the Kohn-Sham eigenvalues and eigenfunctions of momentum $k$ and band $j$ obtained from the scalar-relativistic DFT calculation. Unlike the full polarization the constrained polarization in (\ref{eqPr}) excludes processes taking place within the target space (see Fig. \ref{fig:fig1}(b)). 
For the cRPA calculations and the Wannier function transformation described below the Density Response Code~\cite{kozhevnikov_2010} developed for the Elk code is used.

\begin{figure}[h!]
\includegraphics[width=1.0\columnwidth]{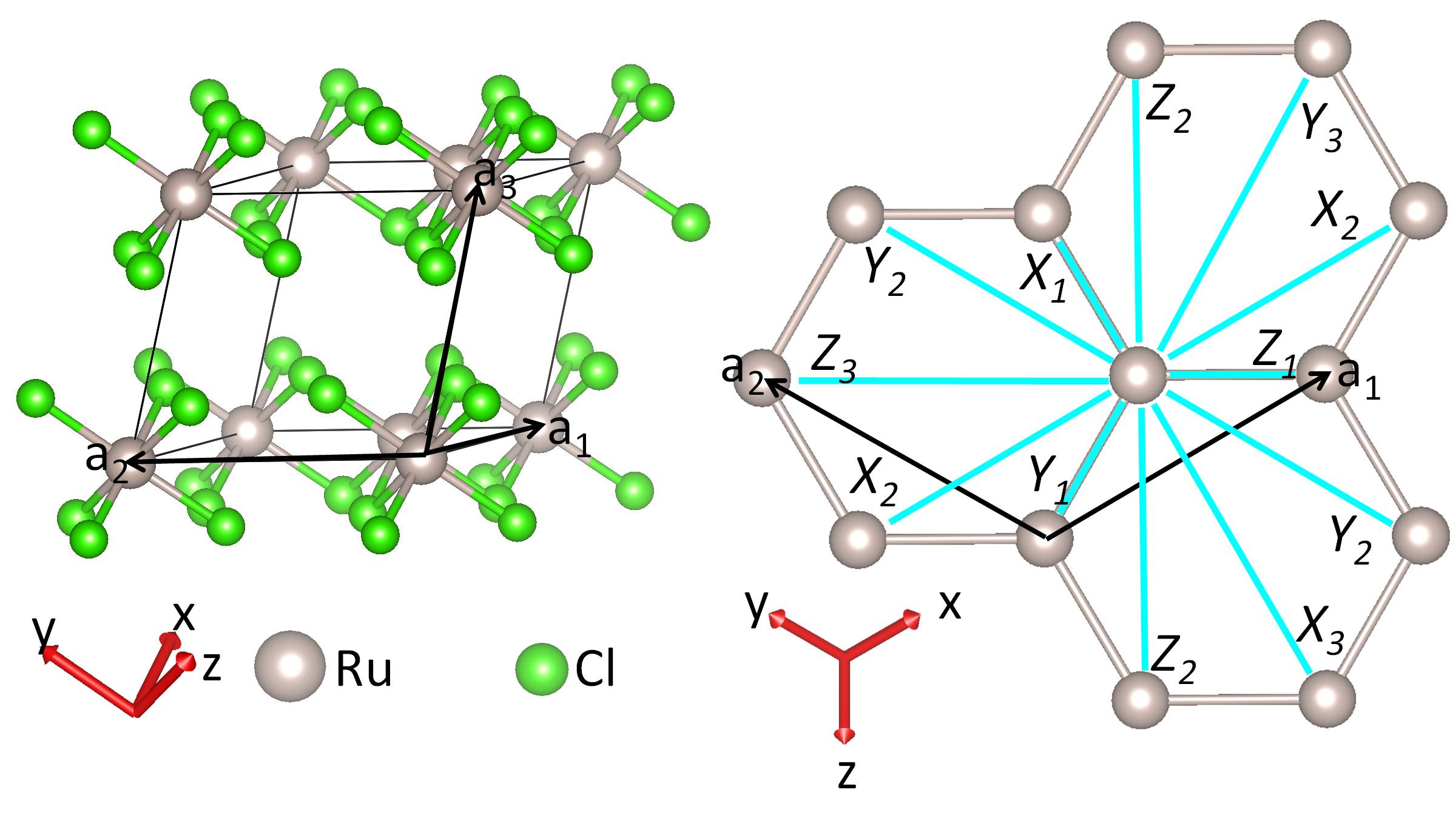}
 \caption{(color online) Definition of the first ($X_1,Y_1,Z_1$), second ($X_{2},Y_{2},Z_{2}$) and third ($X_3,Y_3,Z_3$) nearest neighboring Ru-Ru bonds (cyan lines) and the local coordinates $x,y,z$ (red arrows)  relative to the primitive lattice vectors $a_1,a_2,a_3$ (black arrows) of the $C2/m$ unit cell of $\alpha$-RuCl$_3$.}
 \label{fig:fig2}
\end{figure}

The third step is to apply a transformation of the Kohn Sham eigenfunctions to obtain the generalized multi-orbital Hubbard model in the basis of Ru-$t_{2g}$ Wannier functions.~\cite{marzari_1997} Specifically a projected Wannier function transformation ~\cite{ku_2002,anisimov_2005,kozhevnikov_2010} is performed as follows:
\begin{eqnarray}
|rn\rangle=\sum_{kj}e^{ik\cdot R(r)}|kj\rangle\langle kj|\phi_n\rangle M_{nn'}(k)
\end{eqnarray}
where $r$ and $n$ label the atom and orbital indices of the Wannier states respectively, $R(r)$ labels the unit cell of atom $r$, $|\phi_n>$ are the projected atomic orbital states and $M_{nn'}(k)$ is the L\"owdin orthogonalization matrix~\cite{mayer_2002} given by:
\begin{eqnarray}
M_{nn'}^{-2}(k)=\sum_j \langle \phi_n|kj\rangle\langle kj|\phi_n\rangle
\end{eqnarray}
which ensures the Wannier functions are orthonormal. In our study the projected atomic orbitals $|\phi_n>$ are taken to be the Ru-$t_{2g}$ orbitals $yz,xz,xy$ expressed in the local basis defined in Fig.~\ref{fig:fig2} such that the local coordinates ($x,y,z)$ are approximately along the Ru-Cl bonds and the Ru-$t_{2g}$ satisfy the symmetry properties detailed in Ref. ~\cite{winter_2016}. In the calculations with spin-orbit coupling the spins are rotated into the local coordinate system with the quantization axis along the local $z$ axes. From here we derive the multi-orbital generalized $t_{2g}$ Hubbard model. First we define the scalar relativistic on-site energy and hopping parameters  
\begin{eqnarray}
\varepsilon^{sr}_{n_1n_2}&=&\sum_{kj} \langle r_1 n_1| kj\rangle \epsilon_{kj} \langle kj|r_1n_2\rangle \\
t^{sr,r_2-r_1}_{n_1n_2}&=&\sum_{kj} \langle r_1 n_1| kj\rangle \epsilon_{kj} \langle kj|r_2n_2\rangle 
\end{eqnarray}
with $r_1\ne r_2$ and $\epsilon_{kj}$ and $|kj\rangle$ the Kohn-Sham eigenvalues and eigenstates. Similarly we define on-site energy and hopping parameters from the DFT calculations in which the spin-orbit coupling is included in the second variational treatment. 
\begin{eqnarray}
\varepsilon^{vt}_{n_1\sigma_1n_2\sigma_2}&=&\sum_{kj} \langle r_1 n_1 \sigma_1| k j \rangle \epsilon_{k j} \langle k j|r_1n_2\sigma_2\rangle \\
t^{vt,r_2-r_1}_{n_1\sigma_1n_2\sigma_2}&=&\sum_{k j} \langle r_1 n_1 \sigma_1| k j \rangle\epsilon_{k j} \langle k j |r_2n_2 \sigma_2\rangle  
\end{eqnarray}
From here we define the crystal-field Hamiltonian
\begin{eqnarray}\label{Hcf}
H_{cf}=\sum_{r}\sum_{n_1, n_2}\sum_{\sigma}\varepsilon^{sr}_{n_1n_2} c^\dagger_{rn1\sigma}c_{rn2\sigma} +h.c. ,
\end{eqnarray}
the hopping Hamiltonian
\begin{eqnarray}\label{Hhop}
H_{hop}=\sum_{r_1\ne r_2}\sum_{n_1,n_2}\sum_{\sigma}t^{sr,r_2-r_1}_{n_1n_2}c^\dagger_{r_1n_1\sigma}c_{r_2n_2\sigma} +h.c. ,
\end{eqnarray}
the local spin-orbit coupling Hamiltonian
\begin{eqnarray}\label{Hsocloc}
H^{loc}_{soc}&=& 
\sum_{r}\sum_{n_1,n_2}\sum_{\sigma_1,\sigma_2}
\Big(\varepsilon^{vt}_{n_1\sigma_1n_2\sigma_2}-\varepsilon^{sr}_{n_1n_2}\delta_{\sigma_1\sigma_2}\Big)
\nonumber \\ &&
 \Big( c^\dagger_{rn_1\sigma_1}c_{rn_2\sigma_2} +h.c.\Big) ,
\end{eqnarray}
and the non-local spin-orbit coupling Hamiltonian
\begin{eqnarray}\label{Hsocnloc}
 H^{nloc}_{soc}&=& 
\sum_{r_1\ne r_2}\sum_{n_1,n_2}\sum_{\sigma_1,\sigma_2}
\Big(t^{vt,r_2-r_1}_{n_1\sigma_1n_2\sigma_2}-t^{sr,r_2-r_1}_{n_1n_2}\delta_{\sigma_1\sigma_2}\Big)
\nonumber \\
&&  \Big(c^\dagger_{rn_1\sigma_1}c_{rn_2\sigma_2} +h.c.\Big) .
\end{eqnarray}
We restrict the interactions to the local ($r_1=r_2$) and non-local ($r_1\ne r_2$) Hubbard matrices
\begin{eqnarray}\label{eq:Umat}
U(r_1-r_2,n_1,n_2)= 
\lim_{\omega\rightarrow 0}\int d^3x\int d^3x' 
\nonumber \\ 
W(x,x',\omega) \langle r_1 n_1|x\rangle\langle x|r_1n_1\rangle \langle r_2 n_2|x'\rangle\langle x'|r_2n_2\rangle 
\end{eqnarray}
and the local exchange matrix 
\begin{eqnarray}\label{eq:Jmat}
J(n_1,n_2)= 
\lim_{\omega\rightarrow 0}\int d^3x\int d^3x' 
\nonumber \\ 
W(x,x',\omega) \langle r n_1|x\rangle\langle x|r n_2\rangle \langle r n_2|x'\rangle\langle x'|r n_1\rangle 
\end{eqnarray}
From here we obtain the local interacting Hamiltonian
\begin{eqnarray}\label{Hintloc}
&& H_{int}^{loc}=U\sum_{rn} n_{rn\uparrow} n_{rn\downarrow}
+U'\sum_{r,n\ne n'} n_{rn\uparrow} n_{rn'\downarrow}
\nonumber \\ &&
+J_H\sum_{r,n\ne n'}
\left(
c^\dagger_{rn\uparrow}c^\dagger_{rn\downarrow}c_{rn'\downarrow}c_{rn'\uparrow}
-c^\dagger_{rn\uparrow}c_{rn\downarrow}c^\dagger_{rn'\downarrow}c_{rn'\uparrow}
\right)
\nonumber \\ &&
+(U'-J) \sum_{r,n< n',\sigma} n_{r n\sigma} n_{r n'\sigma}
\end{eqnarray}
with $U$ and $U'$ the intra- and inter-orbital Coulomb repulsion and $J_H$ the Hund's coupling. The non-local interacting Hamiltonian is given by
\begin{eqnarray}\label{Hintnloc}
H_{int}^{nloc}=\sum_{m=1}^3\sum_{\langle r,r'\rangle^m}\sum_{n,n'} \sum_{\sigma,\sigma'} V^m n_{r n\sigma} n_{r' n'\sigma'}
\end{eqnarray}
with $\langle r,r'\rangle^m$ denoting $r$ and $r'$ being $m$-th nearest in-plane neighbors and $V^m$ the in-plane $m$-th nearest neighboring Coulomb repulsion.
The $U$, $U'$, $J_H$ and $V^m$ parameters are obtained from orbital averaging the Hubbard and exchange matrices in Eq. (\ref{eq:Umat}) and (\ref{eq:Jmat}). 
After this the multi-orbital Hubbard model is assembled 
\begin{eqnarray}
H_{t_{2g}}=H_{cf}+H_{hop}+H^{loc}_{soc}+H^{nloc}_{soc}+H^{loc}_{int}+H^{nloc}_{int}
\end{eqnarray}

In the last step perturbation theory in the strong coupling limit is performed. To this end the multi-orbital Hubbard model is split in two pieces: the unperturbed part $H_0=H_{cf}+H^{loc}_{soc}+H^{loc}_{int}+H^{nloc}_{int}$ and the perturbation $\Delta=H_{hop}+H^{nloc}_{soc}$. Then $H_0$ is diagonalized exactly and $\Delta$ is treated with second order perturbation theory in the strong coupling limit: 
\begin{eqnarray}\label{eq:2opt}
\langle l| H_{spin} | l' \rangle = \langle l | \Delta \sum_{h} \frac{|h\rangle\langle h|}{E_h-E_l} \Delta |l'\rangle
\end{eqnarray}  
where $|l\rangle$ and $E_l$ are the degenerate low-energy eigenstates and energies of $H_0$ that contain 1 hole in each Ru atom and $|h\rangle$ and $E_h$ are all the high-energy eigenstates and energies of $H_0$ that contain different distributions of the holes. 

To simplify the analysis the states $|l\rangle$ are restricted to the lowest energy Kramers doublet states that are separated from higher energy states by 165 meV or more. With vanishing crystal field these doublet states reduce to the so-called $j_{\rm eff}=1/2$ states.~\cite{winter_2016} 
Because of the $SU(2)$ symmetry within the Kramers doublet states, any linear combination between  the two Kramers states is also a ground state of $H_0$ and, therefore,  the explicit form of  $H_{spin}$ depends on the choice of ``gauge''. To fix the gauge of these doublet states, we first define up and down pseudo-spins in the Kramers doublet as being the states that diagonalize $L_z - S_z$. Then, the overall phase is fixed so that the coefficient of $c^\dag_{r \, yz \uparrow}c^\dag_{r \, yz \downarrow}c^\dag_{r \, zx \uparrow}c^\dag_{r \, zx \downarrow}c^\dag_{r \, xy -\sigma} |0 \rangle$ becomes a real number, where $|0\rangle$ is the vacuum state. We found that this choice of gauge gives more symmetric interactions with respect to the permutation of $X$, $Y$ and $Z$ bonds than the one that diagonalizes $S_z$ used in Ref.~\cite{yamaji_2014_prl}.  We note that the crystal field splits the $j_{\rm eff}=3/2$ excited quartet in two Kramers doublets. The gap between the lowest energy Kramers doublet and the first excited Kramers doublet is 165 meV. The gap between the lowest energy Kramers doublet and the second excited Kramers doublet is 190 meV.
The non-zero matrix elements of $\langle l| H_{spin} | l' \rangle$ are limited to those in which pseudo spins on all sites of states $|l\rangle$ and $|l'\rangle$ are the same except for a pair of Ru sites $r$ and $r'$ connected by the perturbation $\Delta$. This allows us to compactly rewrite the spin Hamiltonian in terms of spin-operators
\begin{eqnarray}\label{eq:hspin}
H_{spin}  = \sum_{m=1}^3\sum_{\langle rr' \rangle^m} \mathbf{S}_{r}\cdot \mathbf{J}_{rr'} \cdot \mathbf{S}_{r'}
\end{eqnarray}  
with $\langle r,r'\rangle^m$ denoting $r$ and $r'$ being $m$-th nearest in-plane neighbors. Other interactions are ignored in this study. Due to the symmetry of the $C2/m$ space group the matrix form of $\mathbf{J}_{rr'}$ is given by 
\begin{eqnarray}\label{eq:JX}
\left[
\begin{array}{ccc}
J^{xy}_m + K^{xy}_m & \Gamma'^{xy}_m+\zeta_m & \Gamma'^{xy}_m-\zeta_m \\
\Gamma'^{xy}_m+\zeta_m & J^{xy}_m +\xi_m & \Gamma^{xy}_m \\
\Gamma'^{xy}_m-\zeta_m & \Gamma^{xy}_m & J^{xy}_m-\xi_m
\end{array}
\right],
\end{eqnarray}
for the $X_m$ bond, 
\begin{eqnarray}\label{eq:JY}
\left[
\begin{array}{ccc}
J^{xy}_m + \xi_m & \Gamma'^{xy}_m+\zeta_m & \Gamma^{xy}_m \\
\Gamma'^{xy}_m+\zeta_m & J^{xy}_m +K^{xy}_m & \Gamma'^{xy}_m-\zeta_m \\
\Gamma^{xy}_m & \Gamma'^{xy}_m-\zeta_m & J^{xy}_m-\xi_m
\end{array}
\right],
\end{eqnarray}
for the $Y_m$ bond and 
\begin{eqnarray}\label{eq:JZ}
\left[
\begin{array}{ccc}
J^{z}_m  & \Gamma^{z}_m & \Gamma'^{z}_m \\
\Gamma^{z}_m & J^{z}_m & \Gamma'^{z}_m \\
\Gamma'^{z}_m & \Gamma'^{z}_m & J^{z}_m+K^{z}_m
\end{array}
\right]
\end{eqnarray}
for the $Z_m$ bond with $m=1,2,3$ in which Dzyaloshinskii-Moriya interactions have been ignored.~\cite{winter_2016} 

Finally to compare with the available experimental studies we consider the following reduced model 
\begin{eqnarray}\label{eq:hred}
H^{red}_{spin}  &=&\sum_{\langle rr' \rangle^1}\Big( J_1 \mathbf{S}_{r}\cdot \mathbf{S}_{r'}+K_1 S_{r}^{\gamma}S_{r'}^{\gamma}+\Gamma_1 S_{r}^{\alpha}S_{r'}^{\beta}
\nonumber \\ &&
+\Gamma_1 S_{r}^{\beta}S_{r'}^{\alpha} \Big)+J_3 \sum_{\langle rr' \rangle^3}\Big(  \mathbf{S}_{r}\cdot \mathbf{S}_{r'} \Big)
\end{eqnarray}  
in which $\{\alpha,\beta,\gamma \}$ is equal to $\{y,z,x \}$, $\{z,x,y \}$ and $\{x,y,z \}$ for the $X_1$, $Y_1$ and $Z_1$ bonds defined in Fig. ~\ref{fig:fig2} and in which the first neighbor Kitaev, Heisenberg and anisotropy parameters $K_1$, $J_1$ and $\Gamma_1$ and the third neighbor Heisenberg parameter $J_3$ are obtained from bond averaging the results in Eq. (\ref{eq:hspin})-(\ref{eq:JZ}) and setting the rest of the parameters to zero. 

\section{Results}

\begin{figure}[h!]
\includegraphics[width=1.0\columnwidth]{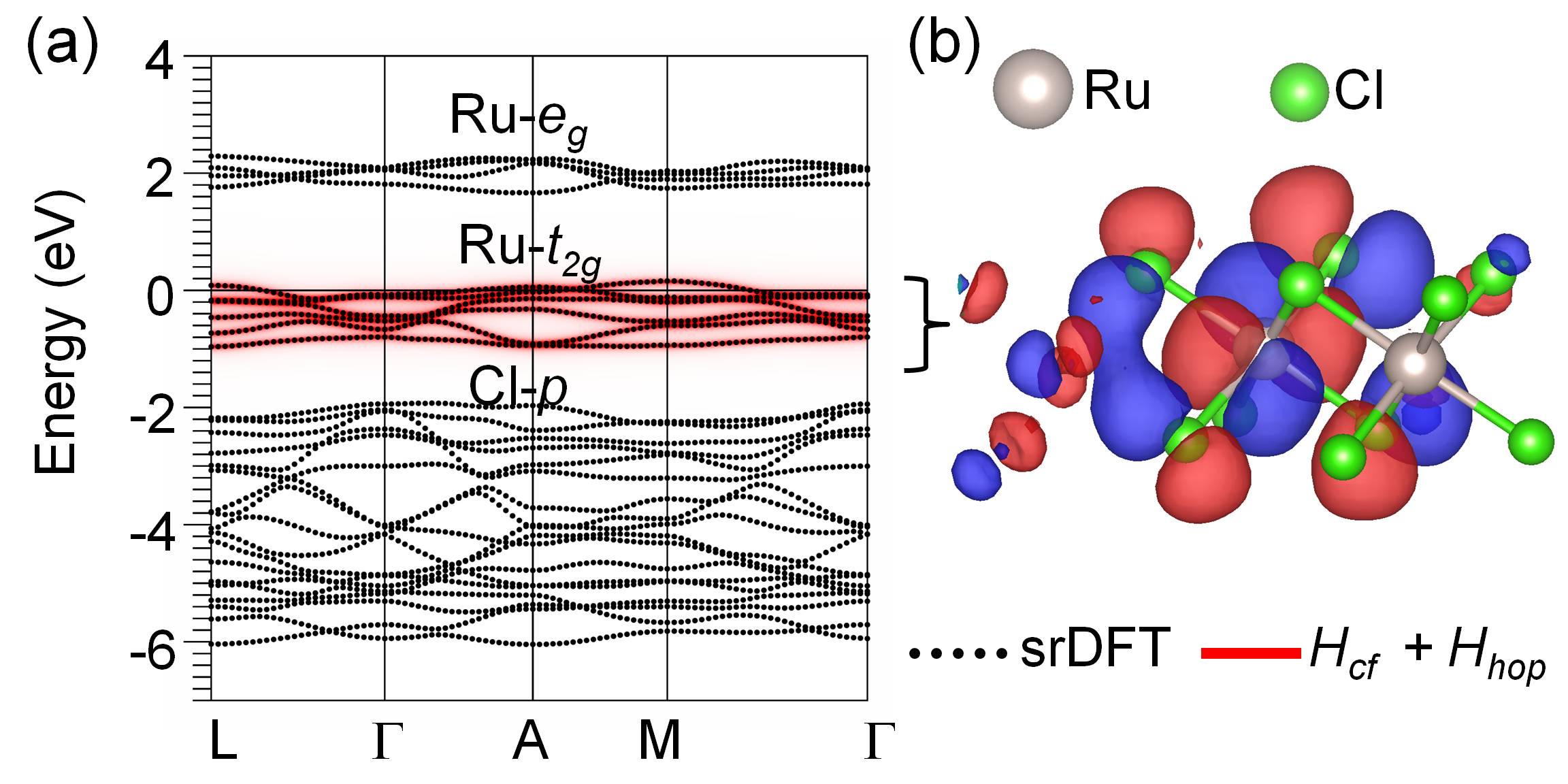}
 \caption{(color online) Left: comparison of the band structure from scalar-relativistic Density Functional Theory (srDFT) and the non-interacting scalar-relativistic part of the Wannier function based Hubbard model: $H_{cf}+H_{hop}$. Right: one of the corresponding Ru-$t_{2g}$ Wannier functions.}
 \label{fig:wannier}
\end{figure}

\begin{table}
\begin{tabular}{llllllllllll}
\multicolumn{1}{c}{} & \multicolumn{3}{c}{local} & \enspace \enspace \enspace & \multicolumn{3}{c}{$Z_1$ bond} & \enspace \enspace \enspace & \multicolumn{3}{c}{$X_1$ bond}  \\
 & $yz$ & $xz$ & $xy$ &  & $yz$ & $xz$ & $xy$ &  & $yz$ & $xz$ & $xy$  \\
\cline{2-4}  \cline{6-8}  \cline{10-12} 
$yz$ & -362 & -7 & -10 &   & 52 & 160 & -21 &    & -98 & -16 & -12 \\
$xz$ & -7 & -362 & -10 &   & 160 & 52 & -21 &    & -16 & 46 &  163 \\
$xy$ & -10 & -10 & -375 &   & -21 & -21 & -150 &    & -12 & 163 & 47  \\
\end{tabular}
\caption{On-site energy and hopping parameters in meV from the non-interacting scalar-relativistic part of the Ru-$t_{2g}$ Wannier function based Hubbard model: $H_{cf}+H_{hop}$.}\label{tab:hop}
\end{table}

In Fig. \ref{fig:wannier} and Tab. \ref{tab:hop} we present results corresponding to the non-interacting part of the generalized Hubbard model without spin-orbit coupling, i.e. $H_{cf}$ and $H_{hop}$
 defined in Eq. (\ref{Hcf}) and (\ref{Hhop}) respectively. Fig. \ref{fig:wannier}(a) shows a comparison of the band structure obtained from scalar relativistic DFT calculation against the one obtained from the non-interacting scalar-relativistic part of the generalized Hubbard model: $H_{cf}+H_{hop}$. Fig \ref{fig:wannier}(b) shows one of the corresponding Wannier functions that displays a $t_{2g}$ character at the center of the Ru atom and strong Cl-$p$ lobes in the nearest neighboring Cl atoms. In Tab. \ref{tab:hop} the local matrix corresponds to $H_{cf}$ on one of the Ru atoms. The hopping matrices correspond to hopping along along the $Z_1$ and $X_1$ bonds defined in Fig.~\ref{fig:fig2}. The crystal field splitting parameters and hopping parameters shown in Tab. \ref{tab:hop} obey the symmetry properties detailed in Ref. ~\cite{winter_2016}. The hopping parameters corresponding to the second and third nearest neighboring in-plane Ru-Ru bonds agree with those reported in Ref. ~\cite{winter_2016} within 1 meV.       

\begin{table}
\begin{tabular}{llllllllllll}
\multicolumn{1}{c}{} & \multicolumn{3}{c}{U local} & \enspace \enspace \enspace & \multicolumn{3}{c}{U $Z_1$ bond} & \enspace \enspace \enspace & \multicolumn{3}{c}{J local}  \\
 & $yz$ & $xz$ & $xy$ &  & $yz$ & $xz$ & $xy$ &  & $yz$ & $xz$ & $xy$  \\
\cline{2-4}  \cline{6-8}  \cline{10-12} 
$yz$ & 2576 & 1895 & 1899 &   & 827 & 893 & 923 &   & & 286 & 288 \\
$xz$ & 1895 & 2576 & 1899 &   & 893 & 827 & 923 &   & 286 & & 288 \\
$xy$ & 1899 & 1899 & 2587 &   & 923 & 923 & 1014 &   & 288 & 288 &  \\
\end{tabular}
\caption{Elements of the local and non-local Hubbard U matrices and local exchange matrix in meV.}\label{tab:int}
\end{table}

Tab. ~\ref{tab:int} shows part of the Hubbard and exchange matrices defined in Eq. (\ref{eq:Umat}) and (\ref{eq:Jmat}). The orbital dependence is relatively weak. Variations are on the order of 10 meV. The orbitally averaged values of the interaction parameters defined in Eq. (\ref{Hintloc}) are given by the intra- and inter-orbital Coulomb repulsions $U=2.58$ eV and $U'=1.9$ eV and the Hund's coupling $J_H=0.29$ eV. The first, second and third nearest neighbor repulsions defined in Eq. (\ref{Hintnloc}) are given by $V_1=0.9$ eV, $V_2=0.53$ eV and $V_3=0.44$ eV respectively. Our interaction parameters derived for $\alpha$-RuCl$_3$ closely resemble the values $U=2.7$ eV $J_H=0.28$ eV and $V_1=1.1$ eV obtained from cRPA calculations for another Ru based compound SrRu$_2$O$_6$~\cite{tian_2015}. We note that in general large non-local Coulomb repulsions are expected in realistic models of materials because of the slow decay of the bare Coulomb potential and the fact that screening within the target space should not be included in the derivation of the model parameters to avoid double counting those effects. ~\cite{aryasetiawan_2004}. For example Hubbard models derived from cRPA for Fe pnictides and chalcogenides ~\cite{mikaye_2010}, ruthenates ~\cite{tian_2015} and iridates ~\cite{yamaji_2014_prl} all display significant non-local Coulomb repulsions relative to their intra-atomic Coulomb repulsions. While the non-local Coulomb repulsions have been ignored in many of the previous derivations of the spin-models for $\alpha$-RuCl$_3$ ~\cite{kim_2015,kim_2016,winter_2016,hou_2017,wei_2017} they have a significant effect on the magnetic interactions as we will discuss below.         

\begin{table*}
\begin{tabular}{lllllllllllllllllllllll}
\multicolumn{1}{c}{} & \multicolumn{6}{c}{atomic-orbital fit $\frac{\lambda}{2} L\cdot S$} & \enspace \enspace \enspace & \multicolumn{6}{c}{local} & \enspace \enspace \enspace & \multicolumn{6}{c}{$Z_1$ bond}   \\
 & $yz\uparrow$ & $xz\uparrow$ & $xy\uparrow$ & $yz\downarrow$ & $xz\downarrow$ & $xy\downarrow$ &  & $yz\uparrow$ & $xz\uparrow$ & $xy\uparrow$ & $yz\downarrow$ & $xz\downarrow$ & $xy\downarrow$ & & $yz\uparrow$ & $xz\uparrow$ & $xy\uparrow$ & $yz\downarrow$ & $xz\downarrow$ & $xy\downarrow$  \\
\cline{2-7}  \cline{9-14}  \cline{16-21} 
$yz\uparrow$ &  & 59$i$ & 0 & 0 & 0 & -59 &    &  & 58$i$ & $i$ & 0& -1-$i$ & -59+$i$&    & 0 &1+2$i$ & 0 & 0 &1+$i$ & 2+12$i$  \\
$xz\uparrow$ & -59$i$ &  & 0 & 0 & 0 & 59$i$ &    & -58$i$ &  & -$i$ & 1+$i$ & 0 & -1+59$i$&    & 1-2$i$ & 0 & 0 & -1-$i$ & 0 & -12-2$i$  \\
$xy\uparrow$ & 0 & 0 &  & 59 & -59$i$ & 0 &    & -$i$ & $i$ &  & 59-$i$ & 1-59$i$ & 0 &    & 0 & 0 & 2 & -2-12$i$ & 12+2$i$ & 0  \\
$yz\downarrow$ & 0 & 0 & 59 &  & -59$i$ & 0 &    & 0 & 1-$i$ & 59+$i$ &  & -58$i$ & -$i$ &    & 0 &-1+$i$ & -2+12$i$ & 0 & 1-2$i$ & 0  \\
$xz\downarrow$ & 0 & 0 & 59$i$ & 59$i$ &  & 0 &    & -1+$i$ & 0 & 1+59$i$ & 58$i$ &  & $i$ &    & 1-$i$ & 0 & 12-2$i$ & 1+2$i$ & 0 & 0 \\
$xy\downarrow$ & -59 & -59$i$ & 0 & 0 & 0 &  &    & -59-$i$ & -1-59$i$ & 0 & $i$ & -$i$ &  &    & 2-12$i$ & -12+2$i$ & 0 & 0 & 0 & 2  \\
\end{tabular}
\caption{Spin-orbit coupling parameters in meV. Local (middle) and non-local (right) parameters derived via first principles Wannier functions compared to (left) atomic-orbital form of the spin-orbit coupling $\frac{\lambda}{2} L\cdot S$ with spin-orbit coupling constant $\lambda$ fitted to the local part of the spin-orbit coupling derived from first principles.}\label{tab:soc}
\end{table*}

Tab. ~\ref{tab:soc} presents the spin-orbit coupling parameters. Specifically the on-site spin-orbit coupling matrix corresponds to $H^{loc}_{soc}$ defined in Eq. (\ref{Hsocloc}). The $Z_1$ spin-orbit coupling matrix is part of $H^{nloc}_{soc}$ defined in Eq. (\ref{Hsocnloc}). We note that in previous derivations of the spin Hamiltonian for $\alpha$-RuCl$_3$ ~\cite{kim_2015,kim_2016,winter_2016,hou_2017,wei_2017} a form of the spin-orbit coupling based on atomic orbitals is assumed. Here we investigate how well that assumption compares with the spin-orbit coupling derived with first principles Wannier functions. The form of the spin-orbit coupling based on atomic $t_{2g}$ orbitals is worked out for example in Ref. \cite{jones_2009} and is denoted $\frac{\lambda}{2} L\cdot S$ in Tab. ~\ref{tab:soc}. By fitting  this form to $H^{loc}_{soc}$ derived from first principles we find the value of the spin-orbit coupling strength to be $\lambda=118$ meV which agrees well with the experimentally reported value of 130 meV reported in Ref. ~\cite{banerjee_2016_nmat}. By comparing the on-site spin-orbit coupling matrix and the atomic orbital fit in Tab. ~\ref{tab:soc} we see that the atomic orbital approximation is nearly perfect for the local part of the spin-orbit coupling. However, we also note that there are significant values of the non-local spin-orbit coupling that are absent in the atomic orbital approximation for the spin-orbit coupling. Specifically there are large non-local spin-orbit couplings between Ru1-$xz$/$yz$ and Ru2-$xy$  orbitals on the order of 12 meV with Ru1 and Ru2 along the nearest neighboring $Z_1$ bond. Similar sized values of the spin-orbit coupling are found along the $X_1$ and $Y_1$ bonds. Along the second and third nearest neighboring bonds the non-local spin-orbit coupling parameters are negligible.  The values of the first neighboring non-local spin-orbit coupling parameters of 12 meV are sizable relative to $\frac{\lambda}{2}=59$ meV given that for each local spin-orbit coupling there are three nearest neighboring non-local spin-orbit couplings on the honeycomb Ru lattice. We note that also in Ref. ~\cite{yamaji_2014_prl} for the closely related compound Na$_2$IrO$_3$ a similar structure of the non-local spin-orbit coupling is reported where the elements between Ir1-$xz$/$yz$ and Ir2-$xy$ orbitals with Ir1 and Ir2 along the $Z_1$ bond are significant relative to $\frac{\lambda}{2}$ in that system. The origin of the non-local spin-orbit couplings in $\alpha$-RuCl$_3$ and Na$_2$IrO$_3$ and in general any transition metal halide, pnictide or chalcogenide is the strong hybridization between the transition metal $d$  orbitals and the anion $p$ orbitals examplified by the Wannier function shown in Fig. \ref{fig:wannier}(b). An interesting question is what the influence of such non-local spin-orbit coupling parameters will be on the magnetic exchanges in $\alpha$-RuCl$_3$.

\begin{table}
\begin{tabular}{lllllllllll}
$m$      & $J_m^{xy}$ & $J_m^{z}$ & $K_m^{xy}$ & $K_m^z$ & $\Gamma_m^{xy}$ & $\Gamma_m^{z}$ & ${\Gamma'}_m^{xy}$ & ${\Gamma'}_m^{z}$  & $\xi_m$ & $\zeta_m$ \\
\hline
$1$      &  -0.7    &  -2.6   & -15.3    & -14.7 & 9.1           &  12.2 & -2.2             & -2.6             &  0.2  &  0.7      \\
$2$      &  0.0     &  0.1    &  -0.6    & -0.8  & 0.0           & 0.0          & -0.1              & -0.1              & 0.0   &  0.0     \\
$3$      &  0.9     &  0.9    &  0.1     & 0.1   & 0.0           & 0.0          & -0.1              & -0.1              & 0.0    & 0.0      \\
\end{tabular}
\caption{Bond-dependent magnetic interaction parameters in meV rounded up to the nearest 0.1 meV for first ($m=1$), second ($m=2$) and third ($m=3$) in-plane nearest neighboring Ru atoms.}\label{tab:bond}
\end{table}

Having obtained the first principles generalized Hubbard model we next perform strong coupling perturbation theory detailed in Eq. (\ref{eq:hspin})-(\ref{eq:JZ}) to derive the magnetic interactions shown in Tab. \ref{tab:bond}.  Just as for example in Ref. ~\cite{winter_2016} we find that some of the parameters display sizable variations depending on the bond directions. This illustrates the complex dependence of the magnetic interactions on the details of the crystal structure and the need for their derivation from first principles. 

\begin{table}
\begin{tabular}{llllll}
																															& $J_1$	& $K_1$	&	$\Gamma_1$	&	$J_3$	&	$C$\\
\hline
This study case 1 full 						& -1.3 	& -15.1 	& 10.1 	& 0.9 	& -19.1  \\
This study case 2 w/o $H^{nloc}_{int}$ 			& -0.2 	& -4.8 	& 3.1 	& 0.7 	& -5.3\\
This study case 3 w/o $H^{nloc}_{soc}$ 		& -1.3 	& -13.3 	& 9.4 	& 1.0 	& -17.3 \\
Inelastic Neutron Scat.  ~\cite{ran_2017} 									& 0 		& -6.8 	& 9.5 				& 0 		& -6.8 \\
Inelastic Neutron Scat. ~\cite{winter_2017}								& -0.5	&	-5		&	2.5					& 0.5 	& -6.5 \\
Thermal Hall Effect ~\cite{cookmeyer_2018,kasahara_2018}				& -0.5	&	-5		&	2.5					& 0.1125& -6.5 \\
THz Spectroscopy~\cite{wu_2018} 															& -0.35 & -2.8 	& 2.4 				& 0.34 	& -3.9 \\
Anisotropic Susceptibility~\cite{lampenkelley_2018}						& n.a. 	& n.a. 	& 29.2 				& n.a. 	& 16.9 \\
Mag. Specific Heat~\cite{suzuki_2018,kubota_2015,do_2017}	&-1.5 	& -24.4  & 5.3 				& 0 		& -29.0  \\ 
\end{tabular}
\caption{Bond averaged magnetic interaction parameters in meV derived for three different cases compared to experimental reports~\cite{ran_2017,winter_2017,wu_2018,lampenkelley_2018,suzuki_2018,kubota_2015,do_2017,cookmeyer_2018,kasahara_2018} with $C=3J_1+K_1$. The parameters deduced from magnetic specific heat data ~\cite{suzuki_2018} have been bond-averaged.  
} \label{tab:red}
\end{table}

\begin{figure}[h!]
\includegraphics[width=0.85\columnwidth]{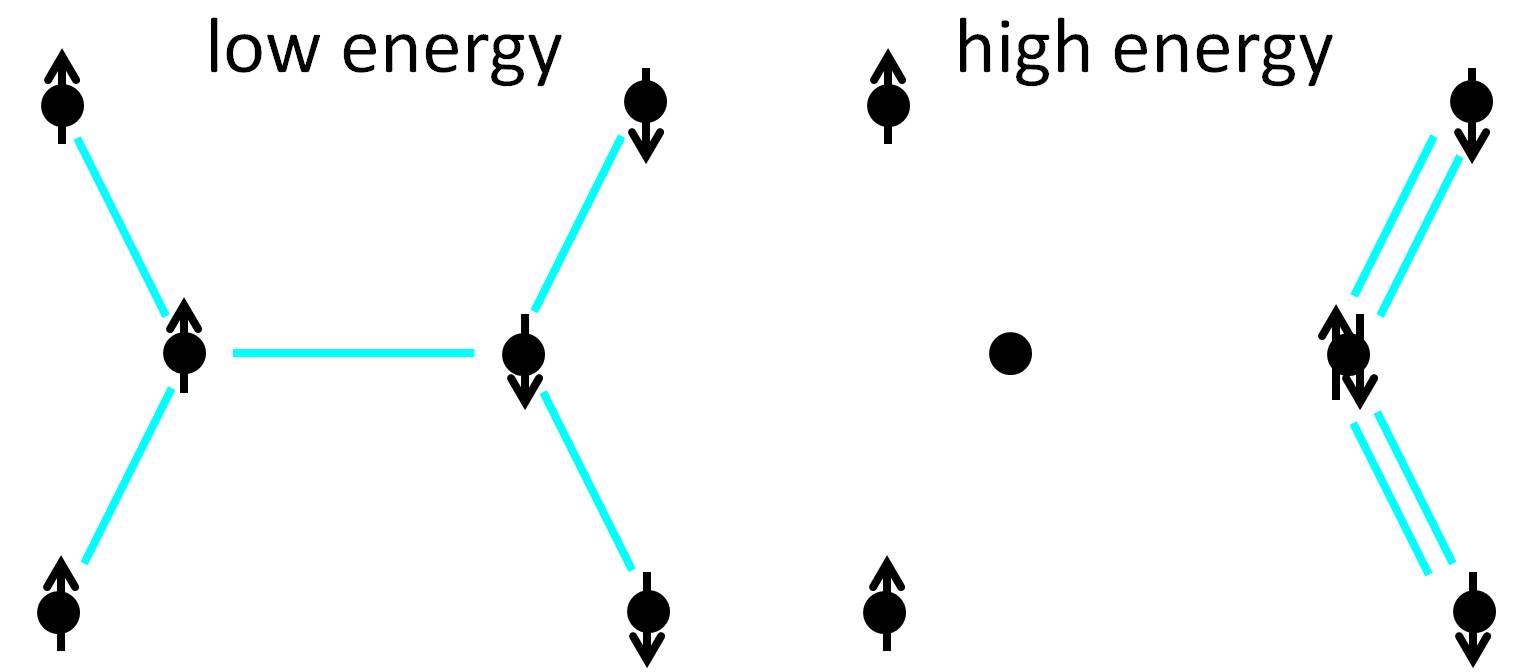}
 \caption{(color online) Comparison of low and high energy state in simplified model with 1 orbital per site. Arrows indicate spin-up and spin-down holes and cyan lines indicate nearest neighboring holes.}
 \label{fig:cartoon}
\end{figure}

Nonetheless we proceed by deriving the parameters of the simplified bond-averaged model defined in Eq. (\ref{eq:hred}) to be able to compare to the available experimental studies. The results are listed in Tab. \ref{tab:red}. Specifically we consider three cases. Case 1 corresponds to the full first principles model. Case 2 and 3 correspond to the first principles model in which the non local interactions and spin-orbit coupling are omitted respectively. When we compare case 1 with case 2 we note that the effect of the non-local interactions is to significantly increase the nearest neighbor magnetic interactions roughly by a factor 3-7. To understand this we consider in Fig. \ref{fig:cartoon} a low and a high energy state in a simple model consisting of 6 sites with 1 orbital per site, nearest neighbor hopping $t$ and local and non-local Coulomb repulsions $U$ and $V$ respectively. By counting nearest neighboring holes in both cases (indicated with cyan lines in Fig.~\ref{fig:cartoon}) we see that the corresponding energies are $E_l=5V$ for the low energy state and $E_h=U+4V$ for the high energy state. When we plug those values into Eq. (\ref{eq:2opt}) we see that the magnetic interactions go as $t^2/(U-V)$ instead of the usual $t^2/U$. In other words the effect of the non-local repulsions will be to enhance the magnetic interactions. These enhancements are quite strong given the large size of the nearest neighbor non-local interaction $V_1=0.9$ eV compared to the intra-atomic repulsion $U=2.58$ eV. From comparing case 1 and 3 we note that the effect of the non-local spin-orbit coupling is not as dramatic as that of the non-local interactions. Still its influence is non-negligible for the first neighbor Kitaev ($K_1$), increasing it by about 14\%. 

\section{Discussion}

From Tab. \ref{tab:red} we see that the parameters derived in our fully first principles study differ significantly compared to the values of previous experimental fits. Why this is the case remains an open question. First of all, it is important to note that there are significant variations in the experimentally derived parameters. For example the Kitaev coupling $K_1$ derived from THz spectroscopy and magnetic specific heat differ more than a factor five. Given this, it is possible that the disagreement between our theoretical results and the experiments reflect difficulties in deducing the parameters from the experiments. Here we will not discuss possible issues on the experimental side and focus on potential problems in the first-principles derivation instead. The derivation of the magnetic parameters in this study is based on unbiased first principles calculations. Nonetheless, approximations are made in the derivation of the low energy electronic Hamiltonian ($H_{t_{2g}}$) and the strong coupling perturbation theory.  

In the strong coupling limit the effects of interactions and spin-orbit coupling form the starting point of the analysis and the hopping parameters are treated as a perturbation. The opposite limit however has also been considered for Na$_2$IrO$_3$~\cite{mazin_2012} and $\alpha$-RuCl$_3$~\cite{johnson_2015} in which the strong anion $p$ assisted hopping between transition metal $t_{2g}$ orbitals leads to the formation of Quasi Molecular Orbitals on the Ru hexagons similar to the molecular orbitals in benzene. Moreover in Ref. ~\cite{winter_2016} it was concluded that the strong coupling perturbation theory for $\alpha$-RuCl$_3$ and A$_2$IrO$_3$ with A=Na,Li does not converge and that instead exact diagonalization is needed to derive the magnetic interactions from the Hubbard model. This conclusion was reached even without taking into account the non-local interactions which push $\alpha$-RuCl$_3$ further away from the strong coupling limit. If indeed $\alpha$-RuCl$_3$ not being in the strong coupling limit is the reason for the large mismatch between the experimental fits and our theoretical results this would be an important point given that many of the theoretical derivations for $\alpha$-RuCl$_3$ or quantum spin liquid materials in general are based on this approximation. It should be noted that in Ref.~\cite{yamaji_2014_prl} the exact same procedure was followed as in this paper to derive the spin model for Na$_2$IrO$_3$ which was found to agree well with experiments. Given that the Ir-$5d$ orbitals are more delocalized compared to the Ru-$4d$ orbitals one would expect that $\alpha$-RuCl$_3$ is closer to the strong-coupling limit than Na$_2$IrO$_3$. One interesting possibility would be to apply the exact diagonalization procedure of Ref.~\cite{winter_2016} to the Hubbard model in this work.

A second approximation made in our derivation is the cRPA. There have been recent  reports that the cRPA tends to over-screen the interactions.~\cite{honerkamp_2018,han_2018,tanado_2019} An underestimate of the strength of screened Coulomb repulsions via cRPA would lead to an overestimate of the magnetic exchanges couplings. We note that two of these studies~\cite{honerkamp_2018,han_2018} are based on simple models that possibly could exaggerate the over-screening effects in cRPA compared to cases with realistic electronic structures. In Ref. ~\cite{tanado_2019} on the other hand the target bands are entangled with the rest of the bands, which could also cause discrepancies between different methods unrelated to over-screening.    

In addition to the potential problem of over-screening, there is the issue that the effects of the Coulomb repulsion are double counted. At the level of DFT the effect of interactions within the target space are already included via the Hartree and exchange-correlation functionals in addition to their treatment in the strong coupling perturbation theory. Furthermore, in our derivation of the magnetic exchange coupling we ignored the frequency dependence of the interaction parameters (c.f. Eq. (\ref{eq:Umat}) and (\ref{eq:Jmat})). A constrained GW approximation has been proposed to remedy the double counting issue and to renormalize the frequency dependence of the interaction parameters into static ones. ~\cite{hirayama_2013,hirayama_2019}   

We like to point out that in Tab. ~\ref{tab:red} there is a relatively good agreement between the model derived without the non-local interactions (case 2) and the fitted parameters obtained from one of the inelastic neutron scattering experiments.~\cite{winter_2016} Therefore it might be tempting to ignore the non-local interactions. However, given that there is no justification why non-local interactions could be ignored we regard this as a coincidence with no physical meaning. 

Another approximation made in our derivation of the magnetic exchange couplings is the orbital averaging of the interaction matrices. This could influence the magnetic interactions, especially given the relatively strong orbital dependence of the non-local interaction matrices (see for example the interaction matrix along the $Z_1$ bond in Tab. \ref{tab:int}). The orbitally symmetric non-local interaction Hamiltonian $H_{int}^{nloc}$ is diagonal in the eigenbasis of the local part of the Hamiltonian: $H_{cf}+H_{soc}^{loc}+H_{int}^{loc}$. This is why we could treat the non-local interaction as part of the unperturbed part of the Hamiltonian. For an orbitally non-symmetric non-local Hamiltonian this no longer works. In principle, it could be possible to treat the orbitally non-symmetric non-local Hamiltonian as part of the perturbation. Such a treatment will be left to explore in future studies.

Finally we note that our model, as most other models in the literature, does not include the out-of-plane magnetic exchange couplings. The reason is that the inter-layer structure of $\alpha$-RuCl$_3$ has been difficult to resolve experimentally. Various inter-layer structures have been reported including the trigonal space group $P3_112$, the rhombohedral space group $R\bar{3}$ and the monoclinic space group $C2/m$ with AB and ABC stackings.~\cite{cao_2016,park_2016, banerjee_2016_nmat,johnson_2015} The difficulty in determining the out-of-plane structure most likely stems from the weak Van der Waals bonding between the RuCl$_3$ layers that allows the layers to easily slide over one another. Related to this, stacking faults have been reported to be present in $\alpha$-RuCl$_3$ which also adds to the difficulty of resolving the out-of-plane structure.~\cite{cao_2016,banerjee_2016_nmat,johnson_2015}  

Our finding contributes to the understanding and design of quantum spin liquid materials.  Per definition first principles models have no free parameters and can thus help constrain the multitude of models that have been proposed for $\alpha$-RuCl$_3$ in the literature.~\cite{winter_2017,wu_2018, lampenkelley_2018,kim_2016,yadav_2016, winter_2016,suzuki_2018, winter_2017_jpcm,ran_2017,wei_2017}  Furthermore, first principles calculations describe the complexity of the full spin model and its dependence on subtle structural distortions without neglecting or bond-averaging the parameters. Ultimately, an accurate first principles method will allow not only for the understanding of the current quantum spin liquid candidate materials but also for predicting how their properties can be optimized by pressure, chemical doping and hetero-structure engineering or how to design new quantum spin liquid materials virtually via high-throughput computations. Laying out potential problems of the theoretical approaches used in this study for the case of $\alpha$-RuCl$_3$ will motivate the search for improved first principles techniques in future efforts to derive first principles models for quantum spin liquid materials. 

\section{Conclusion}
We have derived the magnetic exchange couplings of $\alpha$-RuCl$_3$ via first principles techniques. To this end we utilized the constrained Random Phase Approximation (cRPA) to derive the Ru-$t_{2g}$ Wannier function based generalized Hubbard model to which we applied second order perturbation theory in the limit of the hopping parameters being small compared to the interactions. We have found that the first, second and third nearest neighboring Coulomb repulsions are significant compared to the on-site ones. Furthermore we found sizable elements in the spin-orbit coupling between orbitals on nearest neighboring Ru atoms that are usually ignored in model treatments of the spin-orbit coupling based on atomic orbitals instead of realistic first principles Wannier functions. We have investigated the effect of both the non-local interaction and the non-local spin-orbit coupling on the magnetic exchange couplings. The non-local interactions are found to strongly enhance the magnetic exchange couplings. The non-local spin-orbit coupling overall has a less dramatic effect although it still has a sizable influence on the Kitaev interaction strength. Our full model that includes the influence of both local and non-local interactions and spin-orbit coupling has magnetic exchange couplings that differ from the ones obtained thus far from experimental fits. Highlighting the importance of non-local electron-electron interaction and spin-orbit coupling effects and laying out potential problems in the combined cRPA, Wannier function and strong coupling theory approach in our study contributes to the understanding and virtual engineering of quantum spin liquid candidate materials via first principles calculations.
  
We thank A. Banerjee, S. E. Nagler, G. J. Hal\'asz, D. A. Tennant, A. M. Samarakoon, R. Valent{\'{\i}}, S. M. Winter and D. Mandrus for valuable suggestions and discussions. P.L., S.O. and T.B. acknowledge support from the Scientific Discovery through Advanced Computing (SciDAC) program funded by the U.S. Department of Energy, Office of Science, Advanced Scientific Computing Research and Basic Energy Sciences, Division of Materials Sciences and Engineering. A portion of the work was conducted at the Center for Nanophase Materials Sciences, which is a DOE Office of Science User Facility. DD-OLCF (ORNL) Award ``MAT160'' of Titan supercomputer time is acknowledged with thanks (A.G.E.).

\FloatBarrier

\appendix

\section{Full bond-dependent parameters for case 2 and case 3}

The bond-dependent interactions for the full ab initio model were given in Table~\ref{tab:bond}. Table~\ref{tab:red} also lists bond-averaged parameters for the case of neglected non-local interactions (case 2) and non-local spin-orbit coupling (case 3). For completeness, the full bond-dependent magnetic interaction parameters for these cases are shown in Tab.~\ref{tab:bond:case2} and Tab.~\ref{tab:bond:case3}, respectively.

\begin{table}
	\begin{tabular}{lllllllllll}
		$m$      & $J_m^{xy}$ & $J_m^{z}$ & $K_m^{xy}$ & $K_m^z$ & $\Gamma_m^{xy}$ & $\Gamma_m^{z}$ & ${\Gamma'}_m^{xy}$ & ${\Gamma'}_m^{z}$  & $\xi_m$ & $\zeta_m$ \\
		\hline
		$1$      &  -0.1    &  -0.5   & -4.6    & -4.5 & 2.7           & 3.6         & -0.7             & -0.8             & -0.1  &  0.2      \\
		$2$      &  0.0     &  0.0    &  -0.3    & -0.4  & 0.0           & 0.0          & 0.0              & 0.0              & 0.0   &  0.0     \\
		$3$      &  0.8     &  0.7    &  0.1     & 0.1   & 0.0           & 0.0          & 0.0              & 0.0              & 0.1    & 0.0      \\
	\end{tabular}
	\caption{Bond-dependent magnetic interaction parameters for case 2, neglecting non-local interactions. The parameters are given in meV rounded up to the nearest 0.1 meV for first ($m=1$), second ($m=2$) and third ($m=3$) in-plane nearest neighboring Ru atoms.}\label{tab:bond:case2}
\end{table}

\begin{table}
	\begin{tabular}{lllllllllll}
		$m$      & $J_m^{xy}$ & $J_m^{z}$ & $K_m^{xy}$ & $K_m^z$ & $\Gamma_m^{xy}$ & $\Gamma_m^{z}$ & ${\Gamma'}_m^{xy}$ & ${\Gamma'}_m^{z}$  & $\xi_m$ & $\zeta_m$ \\
		\hline
		$1$      &  -0.7    &  -2.6   & -13.3    & -13.2 & 8.4           & 11.3         & -2.2             & -2.5             &  0.6  &  0.7      \\
		$2$      &  0.0     &  0.0    &  -0.6    & -0.8  & 0.0           & 0.0          & 0.0              & 0.0              & 0.0   &  0.0     \\
		$3$      &  1.0     &  0.9    &  0.1     & 0.1   & 0.0           & 0.0          & 0.0              & 0.0              & 0.0    & 0.0      \\
	\end{tabular}
	\caption{Bond-dependent magnetic interaction parameters for case 3, neglecting non-local spin-orbit coupling. The parameters are given in meV rounded up to the nearest 0.1 meV for first ($m=1$), second ($m=2$) and third ($m=3$) in-plane nearest neighboring Ru atoms.}\label{tab:bond:case3}
\end{table}

\FloatBarrier

\bibliography{RuCl3_Eichstaedt}

\end{document}